\documentclass[english]{article}
\usepackage[T1]{fontenc}
\usepackage[latin9]{inputenc}
\usepackage{color}
\usepackage{textcomp}
\usepackage{amsmath}
\usepackage{amssymb}
\usepackage{esint}

\makeatletter

\newcommand{\noun}[1]{\textsc{#1}}

\makeatother

\usepackage{babel}
\begin{document}
\title{The electromagnetic vacuum field as an essential ingredient of the
quantum-mechanical ontology }
\author{A. M. Cetto and L. de la Peña}

\maketitle
Instituto de Física Universidad Nacional Autónoma de México, 04510,
Mexico City, Mexico
\begin{abstract}
This paper provides elements in support of the random zero-point radiation
field (\noun{zpf}) as an essential ontological ingredient needed to
explain distinctive properties of quantum-mechanical systems. We show
that when an otherwise classical particle is connected to the \noun{zpf},
a drastic, qualitative change in the dynamics takes place, leading
eventually to the quantum dynamics. In particular, we demonstrate
that in parallel with the evolution of the particle canonical variables
into quantum operators satisfying the basic commutator $\left[\hat{x},\hat{p}\right]=i\hbar$,
also the field canonical variables are transformed, giving rise to
the corresponding creation and annihilation operators $\hat{a}^{\dagger},\hat{a}$,
satisfying $\left[\hat{a},\hat{a}^{\dagger}\right]=1$. This allows
for an explanation of quantum features such as quantum fluctuations,
stationary states and transitions, and establishes a natural contact
with (nonrelativistic) quantum electrodynamics.
\end{abstract}

\section{Introduction}

Occasionally, physicists and philosophers of science express their
discomfort with quantum mechanics (\noun{qm}) under the argument that
certain characteristic properties of quantum systems---of atoms in
particular---are not correctly taken into account or explained by
the usual theoretical narrative, or even appear in open contradiction
with the empirical facts. A first example that comes to mind is of
course Einstein, whose abiding malaise with quantum theory is well
known; other prominent examples are the authors of Refs. \cite{Feyn65}-\cite{Primas}.
The pertinence of the criticisms raised cannot be negated, although
for the practicing physicist they are normally of marginal importance. 

On the other hand, an inspection of the current literature readily
reveals the existence of about two dozen different interpretations
of \noun{qm}, some more popular than others, and none of them experimentally
verified \cite{wiki interpr}. How can it be that a fundamental theory
that provides the basis for a most significant part of contemporary
physics, admits such a variety of alternative, even contradictory
interpretations? No serious physicist or philosopher of science in
his five senses would claim to come up with a better interpretation
of Newtonian mechanics or Maxwellian electrodynamics. Reformulations
of a known accepted theory may appear, of course, but fundamental
theories do not accept reinterpretations. Special relativity did not
reinterpret classical mechanics, it extended mechanics to wider domains,
and together with quantum theory helped to precise the extend of its
applicability. We should conclude that such variety of interpretations
signals a crucial underdetermination of quantum mechanics

To understand the origin of so many different visions about the same
fundamental theory, it is convenient to place ourselves in the context
in which \noun{qm} was born. We recall that the quantum formalism---its
excellent mathematical apparatus that we still use today with success---was
born in the absence of a deep understanding of the quantum phenomenology.
The most important physical guides were Bohr's 1913 theory of the
hydrogen atom and the observed atomic spectra with their numerology,
explained in part precisely by Bohr's theory, on one hand, and the
discovery by de Broglie of the wave aspects associated with the motion
of quantum particles, which almost immediately found empirical verification
(alas not explanation) in the form of particle diffraction, on the
other. 

It is against this background that Heisenberg worked on his version
of the theory, the matrix mechanics \cite{Heis25}. Heisenberg discovered
that the quantum particles have an unavoidable random behavior. Being
persuaded of the completeness of his theory, he took this randomness
for an essential, irreducible trait that neither needs nor admits
a deeper explanation, and referred to it simply as \textit{quantum
fluctuations}. Rather than being received with skepticism by the community,
this denomination found extended and immediate acceptance, so much
so that even today people continue to speak of quantum fluctuations
as an innate phenomenon. An example of how far this conviction has
taken ground is that the quantum fluctuations are made responsible
for the formation of galaxies in the early stages of the universe---without
any proof of their early existence. With his \textit{\emph{postulate,}}
Heisenberg assigned to his nascent theory a noncausal and indeterministic
character. And few physicists of the time objected to this view.

A few months after the advent of matrix mechanics appeared Schrödinger's
wave theory. The new theory helped to get a more complete picture
of the quantum world and was well received by many physicists, particularly
those who were not close to the leaders of the competing theory. A
timely contribution of Schrödinger's work was that it gave the recipe
to construct the matrix elements, which Heisenberg's mechanics required
but had no means to calculate. 

On top of this, the description of the atom as an entity that lives
in an abstract, mathematical space---a Hilbert space, a well defined
mathematical structure---gives no indication of what is taking place
in real space-time. A further difficulty is related to the introduction
of 'quantum jumps' between states, which were (and still are, to a
large extent) taken as a capricious quantum trait, not amenable to
further analysis. Strictly speaking, we rely on a powerful formal
description of the atom, with no associated intelligible picture of
it.

Unknowingly, Schrödinger's wave theory implied the introduction of
a new element into the quantum description. The point is that electron
interference patterns are produced by the accumulation of a high number
of point-like events, each one created by a single electron \cite{Tono89}.
A single particle produces an isolated, randomly located bright point
on the detecting screen, the interference---the wave manifestation---becoming
evident only after very many hits. The conclusion---normally one
that goes unnoticed---is that Schrödinger's wave function refers
not to a single particle, but to an ensemble of them. Well interpreted,
the Schrödinger theory is \textit{intrinsically statistical in nature},
and deals with \textit{ensembles} rather than individual particles. 

Nevertheless, the statistical perspective of the quantum phenomenon
was dismissed in general---and adamantly opposed by the Copenhagen
school in particular, which prevails to date under different guises.\footnote{The statistical description of quantum mechanics was proposed and
defended by a few authors, Einstein being the most prominent of them.
For a more recent discussion see \cite{Ball98}.} This opened the door to another infelicitous ingredient, the observer.
The introduction of an \textit{active} character in order to 'explain'
the reduction of the distinctive quantum mixtures to the pure states
observed, added a subjective ingredient to the already odd quantum
scheme.

All in all, such variety of interpretations and re-interpretations
indicates that something of importance is missing in the theory. Having
so many variations indicates that the issue is actually not one of
interpretation, but of an essential incompleteness. The absence of
an appropriate guiding ontological element has turned the physical
situation into a mystery. On several occasions people have considered
that the solution consists in reshaping the formalism, or adding some
hidden variables to the theory, for instance to restore causality,
as is the case with Bohmian mechanics \cite{Bohm52}. But rather than
the kind of incompletenesses considered by Einstein---any statistical
description is incomplete by nature---we are referring to an essential
ingredient that is missing. The point is that whatever is to be added
to the incomplete theoretical framework should be able to address
simultaneously some of its main puzzles, including not just the nature
of quantum fluctuations; atomic stability, quantum transitions, discrete
atomic spectra, wavelike phenomena and the like should find their
natural explanation in a coherent scheme. 

\section{Structure of this paper}

The present paper is one of a series that deals with the development
of stochastic electrodynamics (\noun{sed}) as a physical foundation
for quantum mechanics \cite{TWM63}-\cite{TEQ}. In previous work
we have shown that, by including the zero-point radiation field\noun{
(zpf)}, \noun{sed} discloses the mechanism leading to stationary states
and to\noun{ }the (nonrelativistic) radiative corrections proper of
\noun{qed} \cite{TEQ,NosRMF13}, in addition to offering a possible
explanation for the so-called quantum jumps \cite{Qjumps}, along
with an understanding of the mechanism of entanglement \cite{NosEntangl}
and of the electron spin as an emergent property \cite{NosEspin}.
In the present paper we exhibit a deeper role of the \noun{zpf} in
relation with quantization. At the outset, in section 3 arguments
are given in favor of the \noun{zpf} as an element needed to complete
the quantum-mechanical ontology. The results presented in the rest
of the paper reaffirm these arguments. In section 4, the permanent
action of the background field on an otherwise classical charged particle
is shown to induce a \textit{qualitative change in the dynamics},
by revisiting the physical process that leads to the particle's canonical
variables $x,p$ being replaced by the corresponding quantum operators
$\tilde{x},\hat{p},$ Inspired by this result, in section 5 we disclose
the mechanism of quantization of the radiation field in interaction
with matter and the subsequent emergence of the corresponding creation
and annihilation operators $\hat{a}^{\dagger},\hat{a}$, satisfying
$\left[\hat{a},\hat{a}^{\dagger}\right]=1$, thus establishing contact
with (nonrelativistic) \noun{qed}. As an illustration of the merit
of the present approach, in section 6 we present the derivation of
the (\noun{qed}) formula for the Einstein coefficient for spontaneous
emissions. Some major implications of the results obtained are discussed
along the paper. The final section contains a couple of brief concluding
remarks. 

\section{The missing principle}

Because atomic stability and transitions are electromagnetic processes,
the missing element must be electromagnetic rather than mechanical
in nature. This element should be able to provide at the same time
an answer to the questions of quantum causality and the related fluctuations,
and of the wavelike behaviour of particles. A natural answer is therefore
to consider the stochastic electromagnetic zero-point field (\noun{zpf})---the
field introduced into the quantum world by Planck in 1912 \cite{Planck12}---as
the most suitable candidate. This proposal is precisely what has been
guiding the investigations under the name of \textit{stochastic electrodynamics}
(\noun{sed}). \noun{sed} was initiated more over half a century ago
as a result of the physical intuition of the British physicist Trevor
Marshall \cite{TWM63}, of recent parting. The name itself of the
theory contains a hint about the two elements of the prescription,
electromagnetic and stochastic. The insertion of the \noun{zpf} into
the picture has the additional virtue of bringing the ensuing theory
closer to the successful quantum electrodynamics\noun{ (qed}), as
it contains the vacuum fluctuations \emph{ab initio}. Yet contrary
to \noun{qed}, \noun{sed} considers these fluctuations embodied in
a real, random Maxwellian field that fills the entire space with an
energy $\left(1/2\right)\hbar\omega$ per normal mode.

Further to providing an explanation for the quantum fluctuations,
a most important problem addressed with the introduction of the \noun{zpf}
is the atomic stability---as intuited already at the dawn of quantum
mechanics by Nernst \cite{Nernst16}. The point is that an atomic
electron, like any electric charge in accelerated motion, radiates,
thereby losing energy. Yet the stationary atomic states have constant
energy. The solution is to add a source of energy that can be absorbed
by the atoms at the required rate. \noun{sed} shows this to be the
case: atoms become stable when they live immersed in the \noun{zpf}
\cite{NosRMF13,TEQ}. Moreover, the atomic electrons make transitions
between states; these transitions---conventionally dubbed spontaneous---are
correctly explained by \noun{sed}, as by \noun{qed}; not by \noun{qm},
due to its approximate (nonradiative) nature. 

By serving as a bridge that connects the individual particles of a
system, this field also serves to generate correlations between their
motions even when they do not interact directly, thus giving rise
to an apparently nonlocal behavior, as is manifested e. g. in quantum
entanglement \cite{NosEntangl}.

Coming back to the conceptual, philosophical considerations, we conclude
that the mystery and magic that have accompanied the quantum world
over decades, may be dissolved by considering the presence of the
\noun{zpf }as a \emph{real}, physical field in permanent contact with
matter. Its introduction as an inseparable ingredient of the ontology
of any quantum system allows us to recover determinism (which in the
presence of stochasticity is to be understood as a statistical determinism),
causality, locality and a degree of objectivity and realism. \emph{Full}
realism, i. e., a description of the dynamics of the atom in space-time,
remains a subject for the future.

\section{Quantization of matter; the onset of operators\label{QuM}}

\subsection{The equation of motion of stochastic electrodynamics}

A usual starting point for the analysis of the (nonrelativistic) particle
dynamics in \noun{sed} is the equation of motion, known as Braffort-Marshall
equation, which corresponds to the Abraham-Lorentz equation of electrodynamics
in the dipole approximation \cite{TWM63}, 
\begin{equation}
m\ddot{\boldsymbol{x}}=\boldsymbol{f}(\boldsymbol{x})+m\tau\boldsymbol{\dddot{x}}+e\boldsymbol{E}(t),\label{4-2}
\end{equation}
where $m\tau\boldsymbol{\dddot{x}}$ stands for the radiation reaction
force, with $\tau=2e^{2}/3mc^{3}$ \cite{Dice}. For an electron,
$\tau\approx10^{-23}$ s. $\boldsymbol{E}(t)$ represents the electric
component of the \noun{zpf} taken in the long-wavelength approximation,
with time correlation given in the continuum limit by ($j,k=1,2,3$)
\begin{subequations} \label{EE}
\begin{equation}
\left\langle E_{k}(s)E_{j}(t)\right\rangle =\delta_{kj}\varphi(t-s),\label{4-4}
\end{equation}
with the spectral function
\begin{equation}
\varphi(t-s)=\frac{2\hbar}{3\pi c^{3}}\intop_{0}^{\infty}d\omega\,\omega^{3}\cos\omega(t-s)\label{4-6}
\end{equation}
corresponding to an energy $\hbar\omega/2$ per mode.

\end{subequations}

Following a statistical treatment, these equations have been shown
to lead under certain nonessential restrictions to both the Schrödinger
and the Heisenberg formulation of \noun{qm} \cite{TEQ}. Here we shall
use the same starting point, to address one of the most intriguing
traits of quantum theory, namely how it is that the dynamical variables
become expressed in terms of operators.

\subsection{Statistical dynamic description in the Markov approximation\label{DYN}}

Let us assume that the mechanical system whose motion we want to study,
gets connected to the \noun{zpf} at some instant $t_{o}$. To analyze
the effect of the different forces appearing in Eq. (\ref{4-2}) on
the particle dynamics, we introduce an expansion of $\boldsymbol{x}(t)$
in terms of powers of the electric charge $e$ \cite{FOOP22},
\begin{equation}
\boldsymbol{\boldsymbol{x}}=\boldsymbol{\boldsymbol{x}}^{(0)}+\boldsymbol{\boldsymbol{x}}^{(1)}+\boldsymbol{\boldsymbol{x}}^{(2)}+...=\boldsymbol{\boldsymbol{x}}^{(0)}+\sum_{s=1}\boldsymbol{\boldsymbol{x}}^{(s)},\label{C12}
\end{equation}
where $x_{k}^{(s)}$ stands for the contribution of order $e^{s}$,
$e$ being here the coupling factor of the particle to the field.
(For neutral electromagnetic particles the coupling would depend on
the charge distribution, which may influence the times involved in
the evoution of the dynamics.) By inserting Eq. \ref{C12} in (\ref{4-2})
one obtains the hierarchy (summation over repeated indices is understood)
\begin{equation}
m\ddot{x}_{i}^{(0)}=f_{i}(x^{(0)})+m\tau\dddot{x_{i}}^{(0)},\label{C16a}
\end{equation}

\begin{equation}
m\ddot{x}_{i}^{(1)}=\left.\frac{\partial f_{i}}{\partial x_{j}}\right|_{x^{(0)}}x_{j}^{(1)}+eE_{i}(t),\label{C16b}
\end{equation}

\begin{equation}
m\ddot{x}_{i}^{(2)}-\left.\frac{\partial f_{i}}{\partial x_{j}}\right|_{x^{(0)}}x_{j}^{(2)}=\frac{1}{2}\left.\frac{\partial^{2}f_{i}}{\partial x_{j}\partial x_{k}}\right|_{x^{(0)}}x_{j}^{(1)}x_{k}^{(1)},\label{C16c}
\end{equation}

\[
......
\]
In Eq. \ref{C16a} the third-order time derivative may be replaced
as customary by the approximate first-order expression $\tau$$\left(df_{i}/dt\right)$.
The solution for $x^{(0)}$ is then seen to decay within a time lapse
of the order of the lifetime $\tau_{d}$ determined by the value of
$\left|df_{i}/dt\right|$ (an order-of-magnitude calculation of a
typical dissipation time gives $\tau_{d}\approx10^{-11}s$ for a system
the size of an atom). This means that the deterministic solution of
the homogeneous part of (\ref{4-2}) (i. e., in the absence of the
\noun{zpf)} disappears within a time of order $\tau_{d}$, during
which the system loses the information of its initial conditions. 

Further, from Eq. (\ref{C16b}) it follows that $x_{i}^{(1)}$ is
a purely stochastic variable, describing a \textit{non-decaying} motion
driven by the electric component of the \noun{zpf}, which may be written
in the form
\begin{equation}
x_{i}^{(1)}=e\int_{-\infty}^{t}ds\mathcal{\mathit{\mathcal{G_{\mathit{ik}}\mathit{\mathrm{(}t,s\mathrm{)\mathit{E_{k}(s),}}}}}}\label{C18}
\end{equation}
where the Green function $\mathcal{G}_{\mathit{ij}}\mathit{\mathrm{(}t,s\mathrm{)}}$
is a solution of the equation
\begin{equation}
m\frac{\partial^{2}}{\partial t^{2}}\mathcal{G}_{\mathit{ik}}\mathit{\mathrm{(}t,s\mathrm{)}}=\left.\frac{\partial f_{i}}{\partial x_{l}}\right|_{x^{(0)}}\mathcal{G}_{\mathit{lk}}\mathit{\mathrm{(}t,s\mathrm{),}}\label{C22a}
\end{equation}
with $\mathcal{G}_{\mathit{ik}}\mathit{\mathrm{(}t,t)}=0$ and
\begin{equation}
.\qquad\mathrm{lim}{}_{s\rightarrow t}\frac{\partial}{\partial t}\mathcal{G}_{\mathit{ik}}\mathrm{(}t,s\mathrm{)=\mathit{\frac{\mathrm{1}}{m}\delta_{ik}.}}\label{C22b}
\end{equation}
The solution of Eq. (\ref{C22a}) satisfying these conditions is 
\begin{equation}
\mathcal{G}_{\mathit{ik}}\mathrm{(}t,s)=\left.\frac{\partial x_{i}(t)}{\partial p_{k}(s)}\right|_{x^{(0)}},\label{C23b}
\end{equation}
where the derivative $\partial x_{i}(t)/\partial p_{k}(s)$, a function
of the exact solution of Eq. (\ref{4-2}), is to be calculated to
zero order in $e$, i. e., at $\boldsymbol{x}^{(0)}$. (\ref{C18})
becomes thus \begin{subequations} \label{xipi} xipi
\begin{equation}
x_{i}^{(1)}=e\int_{-\infty}^{t}ds\left.\frac{\partial x_{i}(t)}{\partial p_{k}(s)}\right|_{x^{(0)}}\mathit{E_{k}(s),}\label{C26a}
\end{equation}
and its time derivative gives
\begin{equation}
p_{i}^{(1)}=e\int_{-\infty}^{t}ds\left.\frac{\partial p_{i}(t)}{\partial p_{k}(s)}\right|_{x^{(0)}}\mathit{E_{k}(s).}\label{C26b}
\end{equation}
\end{subequations} These results show that $x_{i}^{(1)}(t)$ and
$p_{i}^{(1)}(t)$ evolve in response to the \noun{zpf,} at specific
rates determined by the external force. 

As is clear from (\ref{C16c}) and the subsequent equations of the
hierarchy, the higher-order solutions $x_{i}^{(r)}$ are also controlled
(although indirectly) by the field, the external force playing now
an accessory role. Consequently, along the evolution of the system
the dynamics undergoes an \emph{irreversible}, \textit{qualitative}
change, from behaving deterministically at $t_{o}$ under the action
of the applied force, to eventually acquiring indeterministic properties
under the control of the \noun{zpf}.

\subsection{Kinematics of the SED system\label{KIN}}

The observation just made that the particle dynamics becomes eventually
controlled by the background field, calls for an analysis of the evolution
of the kinematics used to describe the particle response to the field.
With this purpose, we set out to calculate the Poisson bracket of
the particle's canonical variables
\begin{equation}
\left\{ x_{j}(t),p_{i}(t)\right\} _{xp}=\delta_{ij},\label{K4}
\end{equation}
with $i,j=1,2,3$. In this equation, the Poisson bracket is naturally
calculated with respect to the variables at time $t$. 

The full set of canonical variables includes also the infinite number
of field modes (a semicolon is used for the set of canonical variables,
to avoid confusion with the Poisson bracket),
\begin{equation}
\left\{ q;p\right\} =\left\{ x_{i},q_{\alpha};p_{i},p_{\alpha}\right\} ,\label{K6}
\end{equation}
with $q_{\alpha},p_{\alpha}$ the canonical variables corresponding
to the mode of the radiation field of (angular) frequency $\omega_{\alpha}$.
(A discrete set of frequencies is considered here, for reasons that
will become clear later.) At the initial time $t_{o}$, when particle
and field start to interact, the full set of canonical variables is
given by

\begin{equation}
\left\{ q_{o};p_{o}\right\} =\left\{ x_{io},q_{\alpha o};p_{io},p_{\alpha o}\right\} ,\label{K8}
\end{equation}
with $\text{\ensuremath{x_{io}},\ensuremath{p_{io}}}$ the initial
values of the particle's variables, and $q_{\alpha o},p_{\alpha o}$
corresponding to the original field modes, which are those of the
\noun{zpf} alone. Because the system is Hamiltonian, the set $\left\{ q;p\right\} $
is related to the set $\left\{ q_{o};p_{o}\right\} $ via a canonical
transformation, and therefore the Poisson bracket of any two functions
$f,g$ can be taken with respect to either of them,
\begin{equation}
\left\{ f,g\right\} _{qp}=\left\{ f,g\right\} _{q_{o}p_{o}}\label{K10}
\end{equation}
\[
=\left\{ f,g\right\} _{x_{o}p_{o}}+\left\{ f,g\right\} _{q_{\alpha o}p_{\alpha o}}.
\]
At any time the particle variables $x_{i},p_{j}$ satisfy the condition
\begin{equation}
\left\{ x_{i}(t),p_{j}(t)\right\} _{qp}=\delta_{ij},\label{K12}
\end{equation}
whence according to Eq. (\ref{K10}),
\begin{equation}
\left\{ x_{i}(t),p_{j}(t)\right\} _{x_{o}p_{o}}+\left\{ x_{i}(t),p_{j}(t)\right\} _{q_{\alpha o}p_{\alpha o}}=\delta_{ij}.\label{K14}
\end{equation}

As noted earlier, due to the radiation reaction force, after a time
of order $\tau_{d}$ the particle loses track of its initial conditions
$\boldsymbol{x}_{o},\boldsymbol{p}_{o}$, whence the first term in
Eq. (\ref{K14}) vanishes ,
\begin{equation}
\left\{ x_{i}(t),p_{j}(t)\right\} _{q_{o}p_{o}\:\overrightarrow{t>\tau_{d}}}\left\{ x_{i}(t),p_{j}(t)\right\} _{q_{\alpha o}p_{\alpha o}}.\label{K16}
\end{equation}
This means that eventually, the particle kinematics becomes defined
by the canonical variables of the field modes with which it interacts,
\begin{equation}
\left\{ x_{i}(t),p_{j}(t)\right\} _{q_{\alpha o}p_{\alpha o}}=\delta_{ij}\qquad(t>\tau_{d}).\label{K18}
\end{equation}
Let us now introduce the normal field variables 
\begin{equation}
a_{\alpha}=e^{i\phi_{\alpha}},\;a_{\alpha}^{*}=e^{-i\phi_{\alpha}},\label{K20}
\end{equation}
with $\phi_{\alpha}$ statistically independent random phases in $\left(-\pi,\pi\right)$,
related to the canonical variables by the transformation rules
\[
\omega_{\alpha}q_{\alpha}^{o}=\sqrt{\hbar\omega_{\alpha}/2}(a_{\alpha}+a_{\alpha}^{*}),\;p_{\alpha}^{o}=-i\sqrt{\hbar\omega_{\alpha}/2}(a_{\alpha}-a_{\alpha}^{*}).
\]
as corresponds to the \noun{zpf}, and define the bilinear form $\left[f,g\right]$
in general as the transformed Poisson bracket with respect to the
new variables, 
\begin{equation}
\left[f,g\right]\equiv\sum_{\alpha}\left(\frac{\partial f}{\partial a}_{\alpha}\frac{\partial g}{\partial a_{\alpha}^{*}}-\frac{\partial g}{\partial a}_{\alpha}\frac{\partial f}{\partial a_{\alpha}^{*}}\right)=i\hbar\left\{ f,g\right\} _{q_{\alpha o}p_{\alpha o}}.\label{K26}
\end{equation}
Applied to the canonical particle variables, this gives
\begin{equation}
\left[x_{i},p_{j}\right]=i\hbar\left\{ x_{i},p_{j}\right\} _{q_{\alpha o}p_{\alpha o}}.\label{K24}
\end{equation}

According to Eq. (\ref{K18}), for times $t>\tau_{d}$ this bilinear
form satisfies the condition
\begin{equation}
\left[x_{i},p_{j}\right]=i\hbar\delta_{ij}.\qquad(t>\tau_{d})\label{K28}
\end{equation}
This is a noteworthy result: it indicates that the symplectic relation
between the particle variables $x_{i}$ and $p_{j}$ becomes eventually
determined by their functional dependence on the normal \noun{zpf}
variables, with the scale given by Planck's constant. In the following
section we discuss this drastic change in more detail.

\subsection{Disclosing the origin of the quantum operators}

Let us consider that sufficient time has elapsed for the system to
have reached the regime in which the electron is in a stationary state
of motion; this is what we call the quantum regime \cite{TEQ}. In
the absence of external radiation, the field is in its ground state,
and correspondingly also the particle is in its ground state, characterized
by an energy $\mathcal{E}_{o}$. In the presence of external excitations
the particle may reach an excited state $n$, with energy $\mathcal{E}_{n}\mathcal{>E}_{o}$. 

Given its universal character, Eq. (\ref{K28}) may be applied to
any state $n$, be it the ground state or an excited state $n$. We
may therefore tag the variables $x_{i},p_{j}$ with the subindex $n$
and write, using (\ref{K26}),
\begin{equation}
\left[x_{i},p_{j}\right]_{nn}=\sum_{\alpha}\left(\frac{\partial x_{n}}{\partial a_{\alpha}}\frac{\partial p_{n}}{\partial a_{\alpha}^{*}}-\frac{\partial p_{n}}{\partial a_{\alpha}}\frac{\partial x_{n}}{\partial a_{\alpha}^{*}}\right)=i\hbar\delta_{ij}.\label{T2}
\end{equation}
In what follows we limit the discussion to the one-dimensional case,
for simplicity. The constant value of this bilinear form implies that
the variables $x_{n}$ and $p_{n}=m\dot{x}_{n}$ are linear functions
of the set $\left\{ a_{\alpha},a_{\alpha}^{*}\right\} $. The field
modes involved are those to which the particle responds, namely those
that can take the particle from state $n$ to another state, say $k.$
We therefore write \cite{QSMF21} (see also \cite{Dice}, Ch. 10),
\begin{equation}
x_{n}(t)=\sum_{k}x_{nk}a_{nk}e^{-i\omega_{kn}t}\mathrm{+c.c.,\qquad\;}p_{n}(t)=\sum_{k}p_{nk}a_{nk}e^{-i\omega_{kn}t}\mathrm{+c.c.,}\label{T4}
\end{equation}
where $a_{nk}$ is the normal variable associated with the field mode
that connects state $n$ with state $k$, and $x_{nk}$, $p_{nk}=-im\omega_{kn}x_{nk}$
are the respective response coefficients. Introduction of these expressions
into Eq. (\ref{T2}) gives 
\[
\left[x,p\right]_{nn}=2im\sum_{k}\omega_{kn}\left|x_{nk}\right|^{2}=i\hbar,
\]
where we recognize the well-known Thomas-Reiche-Kuhn sum rule,
\begin{equation}
\sum_{k}\omega_{kn}\left|x_{nk}\right|^{2}=\hbar/2m.\label{T6}
\end{equation}
Further, since the field variables $a_{nk},a_{n'k}$ connecting different
states $n,n'$ with state $k$ are independent random variables, by
combining Eqs. (\ref{K26}), (\ref{T4}) and (\ref{T6}) one gets
\begin{equation}
\left[x,p\right]_{nn'}=i\hbar\delta_{nn'}.\label{T8}
\end{equation}
The quantities $x_{nk}$ and $a_{nk}$ refer to the transition $n\rightarrow k$
involving the frequency $\omega_{kn}$, whilst $x_{kn}$ and $a_{kn}$
refer to the inverse transition, with $\omega_{nk}=-\omega_{kn}$;
therefore, from (\ref{T4}), $x_{nk}^{*}(\omega_{nk})=x_{kn}(\omega_{kn}),\ p_{nk}^{*}(\omega_{nk})=p_{kn}(\omega_{kn}),\ a_{nk}^{*}(\omega_{nk})=a_{kn}(\omega_{kn}),$
whence Eq. (\ref{T8}) takes the form
\begin{equation}
{\displaystyle \sum_{k}}\left(x_{nk}p_{kn'}-p_{n'k}x_{kn}\right)=i\hbar\delta_{nn\text{\textasciiacute}}.\label{T12}
\end{equation}

Given the structure of this equation, the coefficients $x_{nk}$ and
$p_{nk}$ can be organized as the elements of matrices $\hat{x}$
and $\hat{p}$, respectively, with as many rows and columns as there
are different states, so that (\ref{T12}) becomes
\begin{equation}
\left[\hat{x},\hat{p}\right]_{nn'}=i\hbar\delta_{nn'},\label{T14}
\end{equation}
which is the matrix formula for the quantum commutator
\begin{equation}
\left[\hat{x},\hat{p}\right]=i\hbar.\label{T16}
\end{equation}

We see that as a result of the evolution of the kinematics, the Poisson
bracket of the particle's canonical variables $x,p$ has turned into
the quantum commutator. The relationship between Poisson brackets
and commutators established with great insight by Dirac in the early
days of \noun{qm, }finds herewith a possible physical explanation.

The rest of the formalism is obtained by introducing the vectors representing
the stationary states of the particle on which the operators act,
denoted by $\left|\mathrm{\mathit{n}}\right\rangle $. From this point
on, the Hilbert-space formulation may be completed following the standard
procedure \cite{QSMF21}.

Seen as a component of the operator $\hat{x}$ acting on the system,
$x_{nk}$ represents the response to the field mode $(nk)$ which
takes the system from state $n$ to state $k$ and vice versa, as
expressed by the selection rules for dipolar transitions. This is
illustrated with the calculation of the Einstein coefficient for spontaneous
transitions, presented in section \ref{EB} .

It is clear from the above that the operators $\hat{x},\hat{p}$ do
not describe particle trajectories, and no phase-space description
is associated with them. The Hilbert-space formalism represents a
compact and elegant way of describing transitions between states in
the quantum regime, at the cost of a space-time description of what
happens inside the atom. 

This result offers a response to an intriguing question of \noun{qm}
that sounds as an oxymoron: how is it possible that the description
provided by the quantum formalism of a \textit{stationary} quantum
state is to be made in terms of a collection of \textit{transition}
amplitudes between states? The suggested functionality of the transition
coefficients as the building blocks of the operator representation
has moreover an historical value, since in the hands of Born (and
as was unknowingly suggested by Heisenberg) they became the omnipresent
elements of the quantum description.

\section{Quantization of the field}

\subsection{Describing a field mode in interaction with matter}

We now proceed to develop a description of the effect produced on
the radiation field by its interaction with matter, consistent with
the above. This means that rather than assuming that the (free) radiation
field is quantized \textit{ab initio}, or postulating the canonical
quantization of the field by simple analogy, we shift the focus to
that part of the field that is exchanging energy with \textit{\emph{quantized}}\emph{
}matter in order to understand how the field quantization comes about.
The description to be developed should serve to express any electric
or magnetic component of the interacting field, be it the \noun{zpf}
alone or in combination with an external field. 

Initially, when particle and field got coupled, both particle and
field were 'free' and the entire spectrum of field modes was in principle
involved in the interaction (see however the discussion in section
\ref{V}). However, once matter has become quantized as described
above, the field modes in interaction with it belong to the reduced
set $\left\{ \alpha\right\} =\left\{ nk\right\} $. 

The fact that quantized matter interacts selectively with single field
modes of well-defined frequencies allows us to focus on one such mode,
which amounts to working with a single Fourier component of the field,
having a momentum $\boldsymbol{k}$ with $\left|\boldsymbol{k}\right|=\omega/c$,
polarization $\boldsymbol{\epsilon}$ and angular frequency $\omega=\left|\omega_{nk}\right|$.
The electric and magnetic Fourier components corresponding to this
mode can be expressed, as usual, in terms of the respective canonical
variables $\mathrm{q}(t),\mathrm{p}(t)$ (see, e. g., \cite{CohenT,Mandel}).

Let us assume that the field mode of frequency $\omega$ is in a given
state $\mathrm{n}$. We denote with $\mathrm{q_{n}}(t)$, $\mathrm{p_{n}}(t)$
the canonical variables of this mode. and express these in terms of
a set of normal field variables $a_{\mathrm{nn}'}$, with respective
coefficients $\mathrm{q}_{\mathrm{nn}\text{'}}$, $\mathrm{p}_{\mathrm{nn}'}$:
\begin{equation}
\mathrm{q}_{\mathrm{n}}(t)=\sum_{\mathrm{n}'}\mathrm{q}_{\mathrm{nn}'}a_{\mathrm{nn'}}e^{-i\omega_{\mathrm{n'n}}t}\mathrm{+c.c.,}\qquad\mathrm{p}_{\mathrm{n}}(t)=\sum_{\mathrm{n'}}\mathrm{p}_{\mathrm{nn}'}a_{\mathrm{nn}'}e^{-i\omega_{\mathrm{n'n}}t}\mathrm{+c.c.,}\label{F2}
\end{equation}
where 
\begin{equation}
\mathrm{p}_{\mathrm{nn}'}=-i\omega_{\mathrm{n'n}}\mathrm{q}_{\mathrm{nn}'},\label{F3}
\end{equation}
and $\mathrm{n}'$ denotes the states of the field mode that can be
reached from state $\mathrm{n}$ as a result of its interaction with
matter. Since the field mode has a single well-defined frequency $\omega,$
necessarily $\left|\omega_{\mathrm{n'n}}\right|=\omega$, i. e., $\omega_{\mathrm{n'n}}=\pm\omega$.
Consequently, only two of the coefficients $\mathrm{q}_{\mathrm{nn}'}$
connecting state $\mathrm{n}$ with some other state $\mathrm{n}'$
are different from zero. Since there are no intermediate states, we
may identify the immediately upper state of the field (corresponding
to $\omega_{\mathrm{n'n}}=\omega$) with $\text{\ensuremath{\mathrm{n'=n}}+1}$,
and the immediately lower state (corresponding to $\omega_{\mathrm{n'n}}=-\omega$)
with $\text{\ensuremath{\mathrm{\mathrm{n'}=n}}-1}$, so that Eqs.
(\ref{F2}) become
\begin{equation}
\mathrm{q_{n}}(t)=\mathrm{q}_{\mathrm{nn}+1}a_{\mathrm{nn}+1}e^{-i\omega t}\mathrm{+\mathrm{q}_{nn-1}\mathit{a}_{nn-1}e^{i\omega t}+c.c.,}\label{F4}
\end{equation}
\begin{equation}
\mathrm{p}_{\mathrm{n}}(t)=-i\omega\mathrm{q}_{\mathrm{nn}+1}a_{\mathrm{nn}+1}e^{-i\omega t}+i\omega\mathrm{q}_{\mathrm{nn-1}}\mathit{a}_{\mathrm{nn}-1}e^{i\omega t}+c.c,\label{F6}
\end{equation}
and because of (\ref{F3}),
\begin{equation}
\omega\mathrm{q}_{\mathrm{nn}+1}-i\mathrm{p}_{\mathrm{nn}+1}=0,\qquad\omega\mathrm{q}_{\mathrm{nn}-1}+i\mathrm{p}_{\mathrm{nn}-1}=0.\label{F8}
\end{equation}
In analogy with Eqs. (\ref{T4}), we identify $\mathrm{q}_{\mathrm{nn}+1},\mathrm{q}_{\mathrm{nn}-1}$
with the \textit{active }(or\textit{ response})\textit{ coefficients}
involved in the change of state of the field in interaction with matter,
from n to $\text{\ensuremath{\mathrm{n}+1}}$ and $\mathrm{n}-1$,
respectively.

\subsection{Genesis of the field operators}

Let us now take the Poisson bracket of the canonical field variables
pertaining to the single field mode, 
\begin{equation}
\left\{ \mathrm{q_{n}}(t),\mathrm{p_{n}}(t)\right\} =1,\label{FO2}
\end{equation}
and calculate it with respect to the complete set of canonical variables
at the initial time $t_{0}$. By the same argument used in section
(\ref{KIN}), after a time of order $\tau_{d}$, when the material
part has reached the quantum regime, the system loses track of the
initial values of the particle variables,
\begin{equation}
\left\{ \mathrm{q_{n}}(t),\mathrm{p_{n}}(t)\right\} _{q_{o}p_{o}\:\overrightarrow{t>\tau_{d}}}\left\{ \mathrm{q_{n}}(t),\mathrm{p_{n}}(t)\right\} _{q_{\alpha o}p_{\alpha o}}.\label{K16-1}
\end{equation}
and only the Poisson bracket with respect to the original \noun{zpf}
quadratures $\text{\ensuremath{\mathrm{q}_{\alpha o}},\ensuremath{\mathrm{p}_{\alpha o}} }$
survives,
\begin{equation}
\left\{ \mathrm{q_{n}}(t),\mathrm{p_{n}}(t)\right\} _{\mathrm{q}_{\alpha o}\mathrm{p}_{\alpha o}}=1.\label{F03}
\end{equation}
The similarity transformation from canonical \noun{zpf} quadratures
to normal field variables $a_{\alpha},a_{\alpha}^{*}$, Eqs. (\ref{K18}),
leads to the transformed Poisson bracket, 
\begin{equation}
\left[\mathrm{q_{n}}(t),\mathrm{p_{n}}(t)\right]=i\hbar.\label{FO4}
\end{equation}
On account of (\ref{F4}) and (\ref{F6}), $\mathrm{q_{n}}(t)$ and
$\mathrm{p_{n}}(t)$ depend only on $a_{\mathrm{nn}\pm1}$, and Eq.
(\ref{FO4}) gives therefore
\begin{equation}
{\displaystyle \sum_{\mathrm{n'}}}\left(\mathrm{q}_{\mathrm{nn}'}\mathrm{p^{*}}_{\mathrm{nn}'}-\mathrm{p}_{\mathrm{nn}'}\mathrm{q^{*}}_{\mathrm{nn}'}\right)=i\hbar,\label{FO6}
\end{equation}
with $\mathrm{n'=n}\pm1$. By applying Eq. (\ref{F4}) successively
to $\mathrm{n}$ and to $\mathrm{n}$', we note that 
\begin{equation}
\mathrm{q^{*}}_{\mathrm{nn}'}(\omega_{\mathrm{n'n}})=\mathrm{q}_{\mathrm{n'n}}(\omega_{\mathrm{nn'}}),\;\mathrm{p^{*}}_{\mathrm{nn}'}(\omega_{\mathrm{n'n}})=\mathrm{p}_{\mathrm{n'n}}(\omega_{\mathrm{nn'}}),\;a_{\mathrm{nn}'}^{*}(\omega_{\mathrm{n'n}})=a_{\mathrm{n'n}}(\omega_{\mathrm{nn'}}),\label{F08}
\end{equation}
so that Eq. (\ref{FO6}) can be written in the alternative form
\begin{equation}
{\displaystyle \sum_{\mathrm{\mathrm{n'=n}\pm1}}}\left(\mathrm{q}_{\mathrm{nn}'}\mathrm{p}_{\mathrm{n'n}}-\mathrm{p}_{\mathrm{nn}'}\mathrm{q}_{\mathrm{n'n}}\right)=i\hbar.\label{FO10}
\end{equation}
By identifying $\mathrm{q}_{\mathrm{nn}'}$ and $\mathrm{p}_{\mathrm{n'n}}$
as the elements of matrices $\hat{\mathrm{q}}$ and $\hat{\mathrm{p}}$,
respectively, Eq. (\ref{FO10}) becomes
\begin{equation}
\left[\mathrm{\hat{q}},\mathrm{\hat{p}}\right]=i\hbar,\label{FO12}
\end{equation}
for any state n of the field. 

Equation (\ref{FO12}) tells us that after a time lapse of order $\tau_{_{d}}$,
the\textcolor{red}{\emph{ }}\textcolor{black}{(transformed) Poisson
bracket of the canonical field variables $\mathrm{q}(t),\mathrm{p}(t)$,
has evolved towards the canonical field commutator. The matrix elements
of $\mathrm{\hat{q}},\mathrm{\hat{p}}$ represent the transition coefficients
(or 'amplitudes') between field states.} Notice that, once more, the
Hamiltonian nature of the dynamics ensures the preservation of the
symplectic structure in the process of transition from the initial
Poisson brackets to the final commutators. 

As just stated, the matrix elements of\textcolor{black}{{} $\mathrm{\hat{q}},\mathrm{\hat{p}}$
}determine the possible changes of state of the field mode in interaction
with matter. Since $\mathrm{n'=n}\pm1$, $\hat{\mathrm{q}}$ and $\hat{\mathrm{p}}$
have off-diagonal elements immediately above and below the diagonal
only. Further, from (\ref{F3}) we have $\mathrm{p}_{\mathrm{nn}'}+i\omega_{\mathrm{n'n}}\mathrm{q}_{\mathrm{nn}'}=0$.
Therefore, the normalized matrix $\hat{a}$ and its adjoint, defined
as
\begin{equation}
\hat{a}=\frac{1}{\sqrt{2\hbar\omega}}(\omega\mathrm{\hat{q}}+i\hat{\mathrm{p}}),\;\;\hat{a}^{\dagger}=\frac{1}{\sqrt{2\hbar\omega}}(\omega\mathrm{\hat{q}}-i\hat{\mathrm{p}}),\label{FO14}
\end{equation}
have offdiagonal elements either immediately above or immediately
below the diagonal, meaning that they play the role of annihilation
and creation operators, respectively. In terms of these, (\ref{FO12})
becomes 
\begin{equation}
\left[\hat{a},\hat{a}^{\dagger}\right]=1.\label{FO16}
\end{equation}
We recall that the canonical variables $\mathrm{q_{n}}(t),\mathrm{p_{n}}(t)$,
(\ref{F4}) and (\ref{F6}), describe a field mode of frequency $\omega$
with given momentum and polarization $\left(\boldsymbol{k},\boldsymbol{\epsilon}\right)$.
Since the normal variables $a_{\mathrm{n}\mathrm{n'}}$ pertaining
to different field modes have independent random phases, as indicated
in (\ref{K20}), we may generalize Eq. (\ref{FO16}) to operators
$\hat{a}_{\boldsymbol{k\epsilon}},\hat{a}_{\boldsymbol{k'\epsilon'}}$,
and write
\begin{equation}
\left[\hat{a}_{\boldsymbol{k\epsilon}},\hat{a}_{\boldsymbol{k'\epsilon'}}^{\dagger}\right]=\delta_{\boldsymbol{kk'}}^{3}\delta_{\boldsymbol{\epsilon\epsilon'}}.\label{F017}
\end{equation}
The rest of the quantum formalism is obtained by introducing the vectors
representing the possible states of the field mode on which the operators
act, which are denoted by $\left|\mathrm{n}\right\rangle $, with
$\left|\mathrm{0}\right\rangle $ for the ground state, $\hat{a}\left|\mathrm{0}\right\rangle =0$.
The result of operating with $\hat{a}$ or $\hat{a}^{\dagger}$ on
state $\left|\mathrm{n}\right\rangle $ is obtained by applying (\ref{FO16})
iteratively, as usual:
\begin{equation}
\hat{a}\left|\mathrm{n}\right\rangle =\sqrt{\mathrm{n}}\left|\mathrm{n}-1\right\rangle ,\;\hat{a}^{\dagger}\left|\mathrm{n}\right\rangle =\mathrm{\sqrt{n+1}}\left|\mathrm{n}+1\right\rangle .\label{FO18}
\end{equation}

\subsection{On the meaning of the quantum field operators}

The results obtained above are in agreement with the established fact
that light and matter in interaction exchange radiant energy in well-defined
quantities, $\pm\hbar\omega$. By writing the Hamiltonian operator
associated with a single mode as 
\begin{equation}
\mathrm{\hat{H}}=\frac{1}{2}\left(\omega^{2}\mathrm{\hat{q}^{2}+\hat{\mathrm{p}}^{2}}\right)=\frac{\hbar\omega}{2}\left(\hat{a}\hat{a}^{\dagger}+\hat{a}^{\dagger}\hat{a}\right),\label{D2}
\end{equation}
and using (\ref{FO18}), one gets the well-known expression for the
energy expectation value
\begin{equation}
\mathrm{\left\langle n\right|\hat{H}\left|\mathrm{n}\right\rangle }=\mathcal{E}_{\mathrm{n}}=\left(\mathrm{n}+\frac{1}{2}\right)\hbar\omega,\label{D6}
\end{equation}
whence indeed $\left|\mathcal{E}_{\mathrm{n}}-\mathcal{E}_{\mathrm{n\pm1}}\right|=\hbar\omega$. 

However, given that the effect of the operators $\hat{a},\hat{a}^{\dagger}$
is to lower or raise the number $\mathrm{n}$, respectively, it seems
more appropriate to identify the separate expressions
\begin{equation}
\mathrm{\hat{H}^{a}}=\hbar\omega\hat{a}^{\dagger}\hat{a},\;\mathrm{\hat{H}^{e}}=\hbar\omega\hat{a}\hat{a}^{\dagger},\label{D8}
\end{equation}
with the operators corresponding to the field energy available for
processes involving either absorption or emission of radiation by
matter, respectively, according to the order of the individual operators
acting on state $\left|\mathrm{n}\right\rangle $,
\begin{equation}
\mathrm{\left\langle n\right|\mathrm{\hat{H}^{a}}\left|\mathrm{n}\right\rangle }=\mathrm{n}\hbar\omega,\;\:\mathrm{\left\langle n\right|\mathrm{\hat{H}^{e}}\left|\mathrm{n}\right\rangle }=\mathrm{(n+1)}\hbar\omega.\label{D10}
\end{equation}

The operators $\hat{a},\hat{a}^{\dagger}$ are sometimes called the
\emph{fictitious harmonic oscillators} associated with a given field
mode (see, e. g., \cite{CohenT}). However, the electric and magnetic
components of a field mode are real physical entities; they are described
in terms of the canonical variables $\mathrm{q_{n}}(t),\mathrm{p_{n}}(t)$.
In their turn, the operators $\mathrm{\hat{q}},\mathrm{\hat{p}}$
(or $\hat{a},\hat{a}^{\dagger}$) represent the response of a field
mode to its interaction with quantized matter, which may lead to a
change of state of the field with the concomitant loss, viz gain,
of an energy $\hbar\omega$.

\subsection{On the quantization of the free electromagnetic field}

It is commonly accepted that the description of the radiation field
in terms of operators applies in principle to modes of all frequencies,
and that it is only for practical reasons that a distinction may be
convenient when dealing with different parts of the spectrum. As argued
by Mandel and Wolf \cite{Mandel}, ``in the microwave region of the
spectrum, and still at longer wavelenghts, the number of photons is
usually very large, and we are justified in treating the system classically''. 

This prevalent view stands in contrast with the conclusions derived
from the present work. Equations (\ref{D2})-(\ref{D10}), as all
the previous ones, refer to a field mode that exchanges energy \emph{as
a result of its interaction with quantized matter}. Consequently,
they do not tell much about the (quantum or non-quantum) nature of
the \emph{free} radiation field, or the field emitted or detected
by other means. It is certainly tempting to use the elegant quantum
formalism to describe any radiation field of whatever frequency. After
all, the quantum theory of radiation has gained a special status as
``the most successful and embracing theory of optics'' \cite{Mandel}.
However the conclusions derived from it may not necessarily be applicable
beyond the optical region, or more generally, in cases in which the
field is emitted or detected by a nonquantized source. What is the
evidence that low-frequency waves are quantized? Take for example
a beam (or pulse) of radio waves produced by the oscillating elements
of a transmitting antenna, which in principle can have a continuous
range of energies. When such radiation hits a detector, \textcolor{black}{again
it is received as a pulse. We can only safely say that the radiation
is quantized, when it is produced (or absorbed) as a result of a transition
between quantum states of matter. }

\subsection{A note on the reality of the vacuum field \label{V}}

In quantum theory, the vacuum (or vacuum fluctuations) is conventionally
considered as fictitious, or as a virtual entity created spontaneously
by particle-antiparticle pairs. Without this notion conditioning the
validity of the results presented here, we favor the perspective of
the vacuum as a real Maxwellian electromagnetic field in its ground
state. Precisely because it represents the ground state of the complete
radiation field, no energy can be effectively extracted from it by
means of atomic transitions. as indicated in (\ref{D10}) with $\mathrm{n}=0$;
hence it does not activate optical detectors, and no direct spectroscopic
evidence is available to prove its existence. The fact that it is
not part of the photonic field, however, does neither invalidate its
reality nor diminish the consequences of its permanent action on matter,
as illustrated in the present work. A number of other important results
widely reported in the literature, including the Casimir effect \cite{Milo},
the Van der Waals forces \cite{Boyer}, diamagnetism \cite{Boyer,PeJa}
and the Lamb shift (\cite{Milo}, \cite{TEQ} ch. 6), not to mention
the emergence of characteristic quantum phenomena such as entanglement
and atomic stability \cite{TEQ}, speak to the inescapable reality
of this field.

This automatically leads us to the problem of the infinite energy
content of the \noun{zpf}, when integrated over the entire spectrum,
\begin{equation}
\mathcal{E}=\intop_{0}^{\infty}d\omega\,\rho(\omega)=\frac{\hbar}{2\pi^{2}c^{3}}\intop_{0}^{\infty}d\omega\,\omega^{3}.\label{V2}
\end{equation}
It is not our intention to address here this well-known and not yet
resolved problem that has beset modern physics (see, e. g., \cite{Wein13}),
and for which alternative solutions have been advanced (e. g., \cite{San22}).
Nevertheless, in the light of the perspective gained with the work
on stochastic electrodynamics reported in Refs. \cite{TWM63}-\cite{TEQ}
and the results we have presented here, it is clear that at least
at the level of the present description, Eq. (\ref{V2}) poses no
problem. The capacity of massive particles to respond to the radiation
field is limited by their inertia, meaning that they become transparent
and are unable to respond to infinitely high-frequency field modes
and, for the same reason, they are unable to radiate at such high
frequencies. Further, when already in the quantum regine, atomic matter
interacts with field modes of frequencies lying within a limited range,
as discussed above, and may be considered to be transparent to the
rest of the spectrum. Besides, the cutoff frequency (of the order
of Compton's frequency $mc^{2}/\hbar$) frequently introduced in \noun{qed}
and leading to numerically correct results for the radiative corrections,
signals the existence of a physical limit to the frequencies at which
particles and field modes exchange energy, beyond which other kinds
of phenomena take place. Do physical phenomena still occur at even
higher frequencies? Where does the radiation spectrum end? These are
still open questions, to which cosmologists and high-energy physicists
may eventually find an answer. 

\section{Energy balance; contact with QED\label{EB}}

Let us apply some of the results presented here to a problem of interest,
to get a feeling of the kind of calculations that follow from them.
We are interested in what happens with the average energy of the material
system for $t>t$$_{d}$ , when the particle has lost memory of is
initial conditions and has become controlled by the field. The equation
of evolution for the average energy is obtained by multiplying the
dynamical equation (\ref{4-2}) with $\boldsymbol{p}$ and averaging
over the ensemble; for the system in state $n$, it reads \cite{FOOP22}
\begin{equation}
\frac{d}{dt}\left\langle H\right\rangle _{n}=\tau\left\langle \boldsymbol{p}\cdot\boldsymbol{\dddot{x}}\right\rangle _{n}+\frac{e}{m}\left\langle \boldsymbol{p\cdot}\boldsymbol{E}(t)\right\rangle _{n}.\label{EB2}
\end{equation}
The first term on the r.h.s. represents the average power lost by
radiation reaction, and the second term represents the average power
absorbed by the particle from the field. 

To calculate these terms explicitly we use the results of section
(\ref{QuM}). Specifically, we observe that in Eqs. (\ref{xipi})
the arguments under the integral contain factors of the form $\partial x_{i}(t)/\partial p_{j}(s)$,
$\partial p_{i}(t)/\partial p_{j}(s)$, which may be written in terms
of Poisson brackets,
\begin{equation}
\partial x_{i}(t)/\partial p_{k}(s)=\left\{ x_{k}(s),x_{i}(t)\right\} _{xp},\;\partial p_{i}(t)/\partial p_{k}(s)=\left\{ x_{k}(s),p_{i}(t)\right\} _{xp},\label{C32-1}
\end{equation}
According to what we have learned, for times $s,t$ larger than $\tau_{d}$
these Poisson brackets should be replaced by the corresponding quantum
commutators,
\begin{equation}
\left\{ x_{k}(s),x_{i}(t)\right\} _{xp}\rightarrow\frac{1}{i\hbar}\left[\hat{x}_{k}(s),\hat{x}_{i}(t)\right],\;\left\{ x_{k}(s),p_{i}(t)\right\} _{xp}\rightarrow\frac{1}{i\hbar}\left[\hat{x}_{k}(s),\hat{p}_{i}(t)\right].\label{C34-1}
\end{equation}
 By setting $t_{o}=-\infty$ the initial contribution to the integral
for $t_{o}\leq s\leq t_{o}+\tau_{d}$ may be safely neglected, and
Eqs. (\ref{xipi}) take therefore the form\begin{subequations} \label{xipi^1-1}
\begin{equation}
x_{i}^{(1)}=\frac{e}{i\hbar}\int_{-\infty}^{t}ds\left[\hat{x}_{k}(s),\hat{x}_{i}(t)\right]\mathit{E_{k}(s),}\label{C36a-1}
\end{equation}
\begin{equation}
p_{i}^{(1)}=\frac{e}{i\hbar}\int_{-\infty}^{t}ds\left[\hat{x}_{k}(s),\hat{p}_{i}(t)\right]\mathit{E_{k}(s).}\label{C36b-1}
\end{equation}
\end{subequations}

Writing the commutator in terms of the matrix elements corresponding
to state $n$ with the help of Eqs. (\ref{T4}) and (\ref{T6}) gives
\begin{equation}
\left[\hat{x}_{i}(s),\hat{p_{i}}(t)\right]_{nn}=2im\sum_{i}\sum_{k}\left|x_{ink}\right|^{2}\omega_{ikn}\cos\omega_{ikn}(t-s),\label{EB4}
\end{equation}
whence from Eqs. (\ref{EE}) and (\ref{C36b-1}),
\[
e\left\langle \boldsymbol{p}\cdot\boldsymbol{E}\right\rangle _{n}=
\]
\[
=\frac{4me^{2}}{3\pi c^{3}}\sum_{i}\sum_{k}\left|x_{ink}\right|^{2}\omega_{ikn}\intop_{0}^{\infty}d\omega\,\omega^{3}\int_{-\infty}^{t}ds\cos\omega(t-s)\cos\omega_{ikn}(t-s)
\]
\begin{equation}
=\frac{2me^{2}}{3c^{3}}\sum_{i}\sum_{k}\left|x_{ink}\right|^{2}\omega_{ikn}^{4}\left[\delta(\omega-\omega_{ikn})-\delta(\omega+\omega_{ikn})\right].\label{EB6}
\end{equation}
On the other hand, the average power lost by radiation reaction is
readily calculated using Eqs. (\ref{T4}),
\begin{equation}
\tau\left(\boldsymbol{p}\cdot\boldsymbol{\dddot{x}}\right)_{n}=-\frac{2e^{2}}{3c^{3}}\sum_{i}\sum_{k}\left|x_{ink}\right|^{2}\omega_{ikn}^{4}.\label{EB8}
\end{equation}
With Eqs. (\ref{EB6}) and (\ref{EB8}) introduced in (\ref{EB2})
one gets
\begin{equation}
\frac{d}{dt}\left\langle H\right\rangle _{n}=-\frac{4e^{2}}{3c^{3}}\sum_{i}\sum_{k}^{\omega_{ink}>0}\left|x_{ink}\right|^{2}\omega_{ikn}^{4}.\label{EB10}
\end{equation}

When the system is in its ground state there is no contribution to
the sum in Eq. (\ref{EB10}), which confirms that the energy lost
by radiation is compensated in the mean by the energy extracted from
the \noun{zpf,} and detailed energy balance holds; the ground state
is absolutely stable. By contrast, when the system is in an excited
state, both terms are negative and contribute an equal amount to Eq.
(\ref{EB10}). This gives for the rate of change of the energy (\cite{NosRMF13},
\cite{TEQ} Ch. 6) \begin{subequations}\label{Ank}
\begin{equation}
\frac{d}{dt}\left\langle H\right\rangle _{n}=-\sum_{i}\sum_{k}^{\omega_{ink}>0}\hbar\omega_{ink}A_{ink},\label{EB12a}
\end{equation}
with
\begin{equation}
A_{ink}=\frac{4e^{2}}{3\hbar c^{3}}\sum_{i}\left|x_{ink}\right|^{2}\omega_{ikn}^{3},\label{EB12b}
\end{equation}
\end{subequations} which coincides with the \noun{qed} formula for
the Einstein spontaneous emission coefficient. This serves to demonstrate
that the radiation reaction and the \noun{zpf} contribute equal parts
to the spontaneous emission rate (see \cite{Milo} for a discussion
of this point in the context of \noun{qed}). That the result of the
\noun{sed} calculation is in agreement with \noun{qed} is due to the
fact that---in contrast with \noun{qm}---the radiative terms, including
the \noun{zpf}, are included from the beginning in the description.

\section{Final comments and conclusions}

By recognizing the electromagnetic \noun{zpf} as an essential part
of the quantum ontology, we have addressed several longstanding issues
of quantum mechanics. In particular, its presence assigns a physical
cause to the quantum fluctuations and thus accounts for the (statistical)
indeterministic nature of the quantum phenomenon. The introduction
of the \noun{zpf} is shown to change the (apparently) mechanical \textit{\emph{nature}}
of the quantum problem into an electrodynamic one, and to induce a
drastic, qualitative change in the behavior of an otherwise classical
system. Ironically, the \noun{zpf} itself disappears from the picture
at the level of \noun{qm}, leaving only its ghost behind, in the form
of Planck's constant. 

We have come to identify the \noun{zpf} as a key player. Instead of
merely adding the \noun{zpf} as a recourse to induce a stochastic
behavior in an otherwise classical motion, we have placed the emphasis
on its deeper role, which is to induce a qualitative change of behavior,
leading eventually to \noun{qm}. Because no perturbative calculation
to any order will give rise to a qualitatively different behavior;
the $new$ situation entails a \textit{\emph{shift of }}approach.\footnote{A series of detailed calculations using a perturbative approach to
\noun{sed}, have shown to lead to conclusions inconsistent with \noun{qm};
see, e. g., \cite{ColeH}-\cite{Theo}.} 

By introducing the \noun{zpf} into the picture we are considering
that atomic matter is made of electric charges (or electromagnetic
particles, more generally) that fill the space with radiation as a
result of their permanent wiggling. This does not preclude the possibility
that other fields should be required at some point to complete the
quantum ontology; to arrive at \emph{quantum mechanics,} however,
the electromagnetic vacuum proves to be sufficient. In any case, the
image of an isolated atom or a quantum particle in empty space appears
as an idealization with no real counterpart. If it were experimentally
feasible to introduce an atom in a volume completely free of radiation
including the \noun{zpf}, this could provide a valuable possibility
to subject the theory to testing. Partial results in this regard have
been achieved time ago, by introducing atoms in a cavity that modifies
the distribution of modes of frequencies of atomic interest, and observing
a concomitant effect on the natural decay times of excited states
(for early experimental work, see \cite{Goy83}-\cite{Hulet85}).
As discussed in section \ref{EB}, these decay times are determined
jointly by the radiation reaction \emph{and} the \noun{zpf}. 

The transition from the classical to the quantum regime represents
a most delicate point of the emerging-quantum theory. From the perspective
gained with the present work, one may say that quantum dynamics is
a variant and extension of classical physics into the stochastic domain,
which finds a place of its own, both because of the distinctive behavior
of quantum systems and for the wealth of phenomena and applications
to which it gives rise.

\end{document}